\documentclass[%
 reprint,
 amsmath,amssymb,
 aps,
]{revtex4-2}
\usepackage{graphicx}
\usepackage{dcolumn}
\usepackage{bm}
\usepackage{amsmath,amsfonts,amssymb}
\usepackage{graphicx,epsfig}
\usepackage{color}
\usepackage{captcont}
\usepackage{slashed}
\usepackage{epstopdf}
\usepackage{mathtools}
\usepackage{natbib}
\usepackage{ulem}
\usepackage{subfigure}
\usepackage{graphicx}
\usepackage{CJKutf8}

\newcommand{\be}{\begin{eqnarray}}
\newcommand{\ee}{\end{eqnarray}}
\newcommand{\bea}{\begin{eqnarray}}
\newcommand{\eea}{\end{eqnarray}}

\begin{document}
\begin{CJK}{UTF8}{<font>}
\preprint{APS/123-QED}

\title{High-order QED correction impacts on phase transition of the Euler-Heisenberg AdS black hole}
\author{Guan-Ru Li}
\email{2007301068@st.gxu.edu.cn}
\affiliation{Guangxi Key Laboratory for Relativistic Astrophysics, School of Physical Science and Technology, Guangxi University, Nanning 530004, People's Republic of China}
\author{Sen Guo}
\email{Corresponding author: Sen Guo(sguophys@st.gxu.edu.cn)}
\affiliation{Guangxi Key Laboratory for Relativistic Astrophysics, School of Physical Science and Technology, Guangxi University, Nanning 530004, People's Republic of China}
\author{En-Wei Liang}
\email{Corresponding author: En-Wei Liang(lew@gxu.edu.cn)}
\affiliation{Guangxi Key Laboratory for Relativistic Astrophysics, School of Physical Science and Technology, Guangxi University, Nanning 530004, People's Republic of China}
\date{\today}
\begin{abstract}
Two-phase transition branches of the Euler-Heisenberg (EH) anti-de Sitter (AdS) black hole (BH) were derived from its phase transition critical behavior by Magos $et~al.$ [Phys. Rev. D. 102, 084011 (2020)]. We found that the phase transition is unstable. Considering the high-order quantum electrodynamics (QED) correction, we re-derive the EHAdS BH solution and investigate its critical thermodynamic quantities. It is found that the corrected EHAdS BH has only one stable phase transition branch, and its critical exponents are equivalent to that of the vdW system. From the microscopic point of view, we also derive its normalized scalar curvature based on the Ruppeiner geometry. Different from two concave surfaces of the scalar curvature without considering the high-order QED correction, we show that the corrected Ruppeiner geometry has only one concave surface. Our results indicate that the phase transition instability derived by Magos $et~al.$ is due to without considering the high-order QED correction.
\end{abstract}
\maketitle
\section{Introduction}
\label{intro}
\par
Black hole (BH) thermodynamics is considered as a bridge connecting general relativity, quantum mechanics, and classical thermodynamics. Regarding the BH in the anti-de Sitter (AdS) spacetime as a thermodynamic system, Hawking and Page found that its thermodynamic properties are similar to the classical thermodynamic system \cite{1}. Taking the negative cosmological constant as the thermodynamic pressure, Dolan $et~al.$ developed the AdS BH thermodynamics to the extended phase space \cite{2,3}. By investigating the $P-\upsilon$ critical behavior of the charged AdS BH in the extended phase space, Kubiznak $et~al.$ found that the critical exponents of the charged AdS BH are precisely the same as the van der Waals (vdW) system \cite{4}. Johnson proposed that the charged AdS BH can be modeled as a heat engine as the vdW fluid, and its efficiency can be calculated similarly \cite{5}. Ayd{\i}ner $et~al.$ investigated the Joule-Thomson expansion of the charged AdS BH and obtained the phase transition heating-cooling regions in the $T-P$ plane \cite{6}. Thermodynamics properties of various charged AdS BHs in the extended phase space are also extensively discussed \cite{7,8,9,10,11,12,13,14,15,ye}.

\par
Based on Dirac's positron theory, Euler and Heisenberg proposed a new approach to describe the electromagnetic field. They derived the effective Lagrangian electromagnetic field density by revising Maxwell's equations in the vacuum \cite{16}. This new effective Lagrangian density has the high-order terms of the nonlinear electromagnetic (NEM) field \cite{17}. Within the quantum electrodynamics (QED) framework, Schwinger reformulated this non-perturbative one-loop effective Lagrangian density, which carries the main characteristics of the Euler-Heisenberg (EH) NEM field \cite{18}. If the electric field strength is higher than the critical value ($m^2c^3/e\hbar$), the QED effect leads to the emergence of particle pairs in the vacuum \cite{19}. Coupling the one-loop effective Lagrangian density with the Einstein field equation, Yajima $et~al.$ obtained the EH BH solution \cite{20}. Kruglov provided an approximation approach for the EH NEM field within the high-order QED framework, and gave the corresponding static spherically symmetric BH solution \cite{21}. By utilizing the Newman-Janis algorithm and its Azreg-A\"{\i}nou formulation, Bret\'{o}n $et~al.$ obtained the rotating BH solution in the Einstein EH theory and analyzed its event horizons, ergoregions, and test particle circular orbits \cite{BN-1}. Subsequently, they investigated the birefringence and the quasinormal modes of the spherically symmetric EH BH. The effect of the EH NEM field suppresses the quasinormal modes, making the charged BH behave more Schwarzschild-like \cite{BN}.

\par
Magos $et~al.$ generalized the EH BH solution to the AdS spacetime by considering the negative cosmological constant \cite{22}. They presented that the EHAdS BH is characterized by the BH mass $M$, electric charge $Q$, cosmological constant $\Lambda$, and EH parameter $a$. By deriving the equation of the state and the critical behavior, they found that the critical volume shows two-phase transition branches of the EHAdS BH. The second phase transition branch is similar to the vdW system, and the first phase transition branch depends on the EH NEM field, leading to the phase transition split in a small BH region. However, some AdS BHs has only one phase transition branch and represents similar properties to the vdW system \cite{7,23}. Note that these BH solutions are from the Einstein field equation coupling the NEM fields. We suspect whether the first phase transition branch of the EHAdS BH results from the effect of the high-order QED.

\par
In this analysis, we derive the EHAdS BH correction solution by considering the high-order QED correction and investigate the phase transition of this scenario. We also analyze the critical behavior and equal area law in the cases of the EHAdS BH and the corrected EHAdS BH. It is found that this disturbance correction as a thermodynamic stable probe can be used to reveal the physical properties of the EHAdS BH. The paper is organized as following: In Sec. \ref{sec:2}, we analyze the critical behavior and equal area law in the framework of the EHAdS BH thermodynamics. Our solution of the EHAdS BH with high-order QED correction and investigation of thermodynamic behaviors are presented in Sec. \ref{sec:3}. Our conclusions is presented in Sec. \ref{sec:4}.

\section{The EHAdS BH and thermodynamic characteristics}
\label{sec:2}
\par
\par
The 4-dimensional spherically symmetric line element of the EHAdS BH is given by \cite{22}
\begin{equation}
{\rm d}s^2=-f(r){\rm d}t^2+\frac{1}{f(r)}{\rm d}r^2+r^2 {\rm d}\Omega^{2},
\label{2-0-1}
\end{equation}
where the metric potential $f(r)$ is
\begin{equation}
f(r)=1-\frac{2M}{r}+\frac{Q^2}{r^2}-\frac{aQ^4}{20r^6}-\frac{\Lambda r^2}{3},
\label{2-0-2}
\end{equation}
in which $M$ and $Q$ are the mass parameter and charge of the BH, $a$ is the EH parameter, $\Lambda$ is defined with the thermodynamic pressure $P$, i.e. $\Lambda \equiv -8\pi P$ \cite{2,3}. The BH horizon $r_{+}$ is derived from the largest root of the equation $f(r_{+})=0$. Hence, the mass $M$ can be expressed with the horizon radius $r_{+}$,
\begin{equation}
M=\frac{r_{+}}{2}+\frac{Q^2}{2r_{+}}-\frac{aQ^4}{40r_{+}^5}-\frac{\Lambda r_{+}^3}{6}.
\label{2-0-3}
\end{equation}
The Hawking temperature is given as
\begin{equation}
T=\frac{1}{4\pi r_{+}}\Bigg(1-\frac{Q^2}{r_{+}^2}+\frac{aQ^4}{4r_{+}^6}-\Lambda r_{+}^2\Bigg).
\label{2-0-4}
\end{equation}
Utilizing $\Lambda \equiv -8\pi P$, the state equation of the EHAdS BH from the above equation can be expressed as
\begin{equation}
P=\frac{T}{2r_{+}}-\frac{1}{8\pi r_{+}^2}+\frac{Q^2}{8\pi r_{+}^4}-\frac{aQ^4}{32\pi r_{+}^8}.
\label{2-0-5}
\end{equation}

\par
\begin{figure}[htbp]
  \centering
  \includegraphics[width=7.5cm,height=5cm]{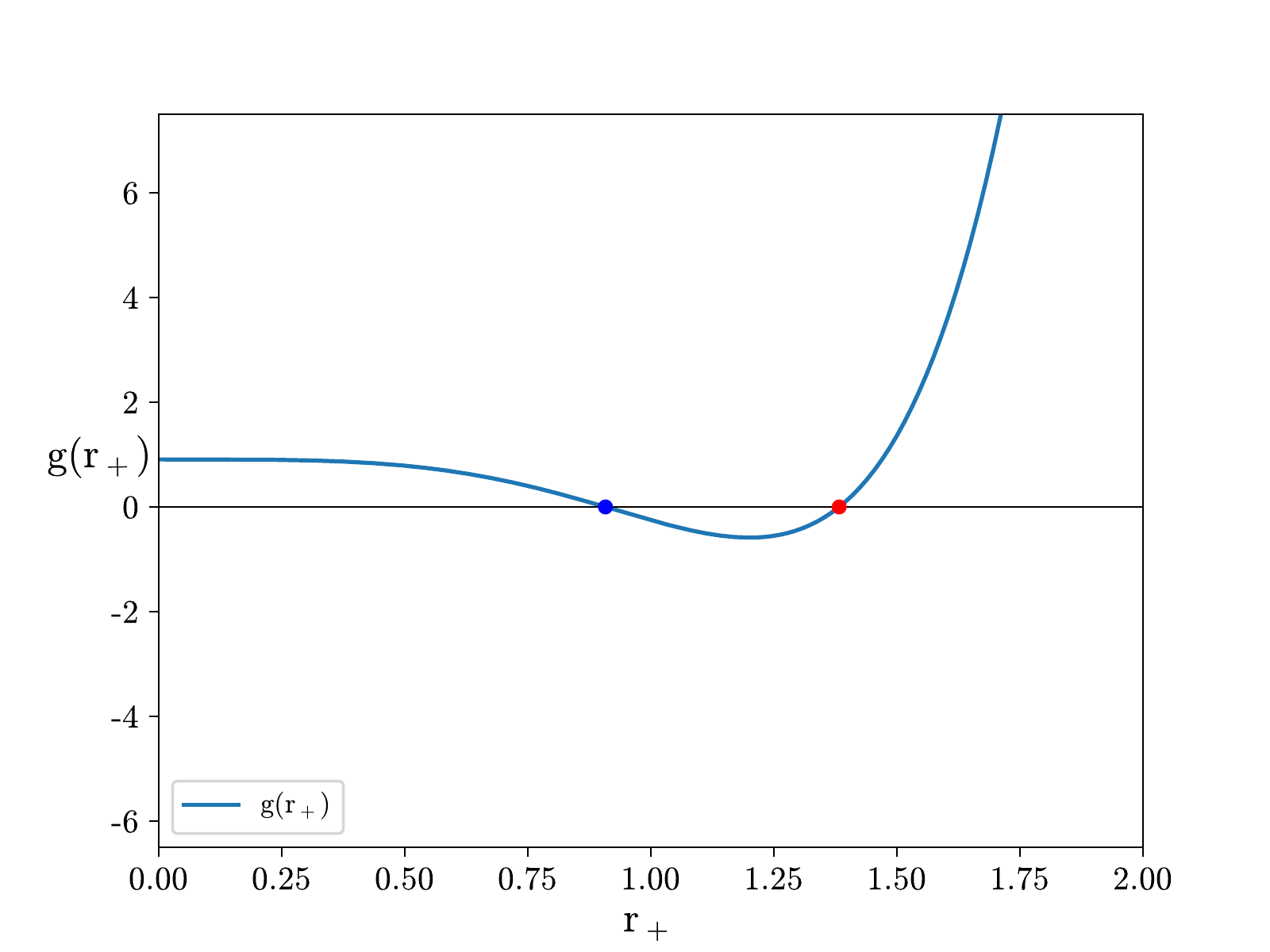}
  \caption {$g({r_{+}})$ as a function of $r_{+}$, where the EH parameter is set as $a=1$ and BH charge is taken as $Q=0.6$}\label{fig:1}
\end{figure}

Based on Eq. (\ref{2-0-5}) and the critical conditions, the critical thermodynamic quantities of the EHAdS BH are obtained, i.e.
\begin{eqnarray}
T_{\rm c}&=&\frac{1}{2\pi r_{\rm c}}\Bigg(1-\frac{2Q^2}{r_{\rm c}^2}+\frac{aQ^4}{r_{\rm c}^6}\Bigg),
\label{2-0-6}\\
P_{\rm c}&=&\frac{1}{8\pi r_{\rm c}^2}\Bigg(1-\frac{3Q^2}{r_{\rm c}^2}+\frac{7aQ^4}{4r_{\rm c}^6}\Bigg),
\label{2-0-7}\\
r_{\rm c}^2&=&2Q^2\Bigg(2\cos\Bigg(\frac{1}{3}\arccos\bigg(1-\frac{7a}{16Q^2}\bigg)-\frac{2\pi k}{3}\Bigg)+1\Bigg),\nonumber\\
&~&~~~~~~~~~~~~~~~~~~~~~~~~~~~~~~~~~~~~~~~~~~~k=0;~1;~2,
\label{2-0-8}
\end{eqnarray}
In case of $k=2$, $r_{\rm c}^2$ is negative. In cases of $k=0$ and $k=1$, they represent the two-phase transition branches respectively. Note that the above critical thermodynamic quantities are derived from the condition of the EH parameter $a$ satisfying $0 \leq a \leq 32Q^2/7$ \cite{22}. Utilizing the critical conditions and state equation, one can obtain
\begin{eqnarray}
g(r_{+})\equiv r_{+}^6-6Q^2r_{+}^4+7aQ^4=0.
\label{2-0-9}
\end{eqnarray}
By setting EH parameter $a=1$ and BH charge $Q=0.6$, $g(r_{+})$ as a function of $r_{+}$ is plotted in Fig.\ref{fig:1}. $g(r_{+})$ function curve intersects the horizontal axis twice (blue point and red point), where the blue point corresponds to the ``first critical point''($k=1$) and red point corresponds to the ``second critical point'' ($k=0$). It indicates that there are two-phase transition branches for the usual EHAdS BH situation.

\par
\begin{figure*}[htbp]
  \centering
  \includegraphics[width=7.5cm,height=5cm]{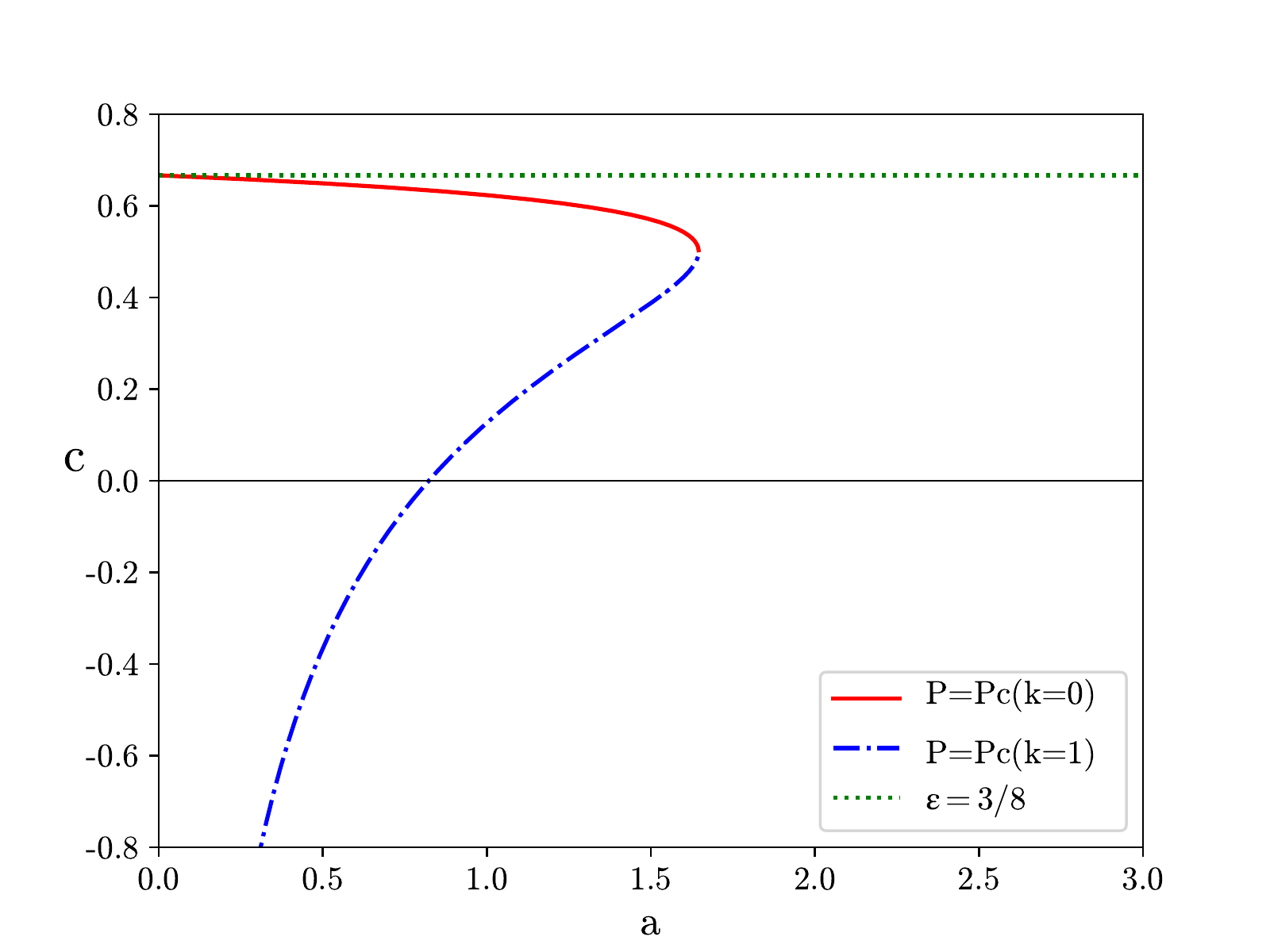}
  \hspace{0.1cm}
  \includegraphics[width=7.5cm,height=5cm]{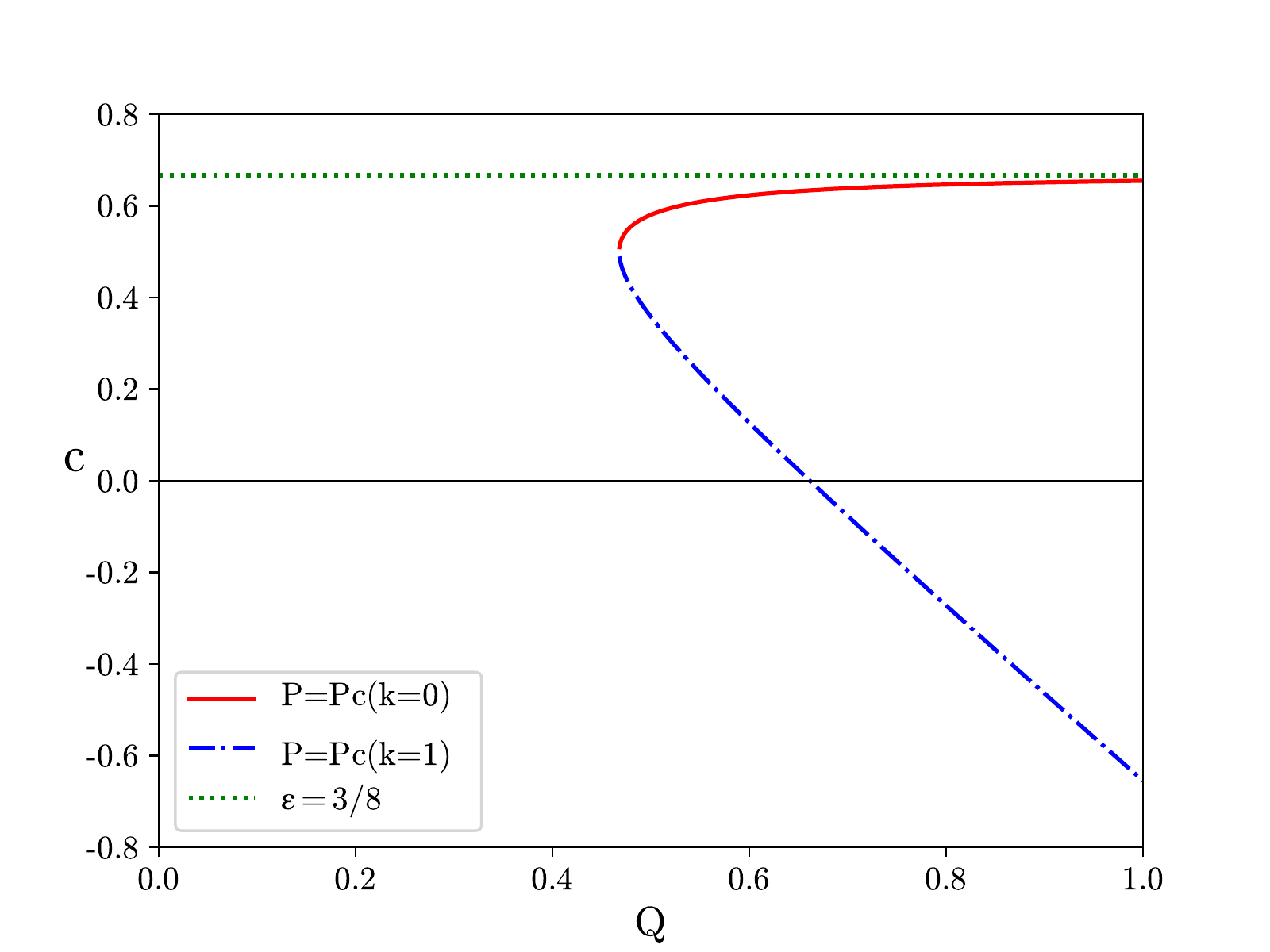}
  \caption {Left panel: $c$ as a function of $a$ for the EHAdS BH with $Q=0.6$. Right panel: $c$ as a function of $Q$ for the EHAdS BH with $a=1$. The red curve is for $k = 0$, the blue curve is for $k = 1$, and the green curve is for $\varepsilon_{\rm vdW} = 3/8$.}\label{fig:2}
\end{figure*}

The universal constant is defined as $\varepsilon \equiv {P_{\rm c}\upsilon_{\rm c}}/{T_{\rm c}}$, where $\upsilon_{\rm c}=2r_{\rm c}$ is the specific volume \cite{4}. Thus, we have
\begin{eqnarray}
\varepsilon = \frac{21c}{16+32c},
\label{2-0-10}
\end{eqnarray}
where $c \equiv 1-2Q^2/r_{\rm c}^2$. If $-1/2 < c < 0$, we have $\varepsilon<0$. Figure \ref{fig:2} shows $c$ as a function of $a$ and $Q$. One can observe that $c$ ranges in $(-\infty,1/2)$ for the $\varepsilon_{\rm 1}$, and $c$ ranges in $(1/2,3/8)$ for the $\varepsilon_{\rm 2}$, where $\varepsilon_{\rm 1/2}$ corresponds to the universal constant of the first/second critical point. Hence, the universal constant at the first critical point could be negative if $8Q^2/7 < a < 16Q^2/7$, which corresponds to unstable configurations of $a$ and $Q$. It is also found that $\varepsilon_{\rm 2}$ is closed to $\varepsilon_{vdW}$.

\par
\begin{figure}[htbp]
  \centering
  \includegraphics[width=7.5cm,height=5cm]{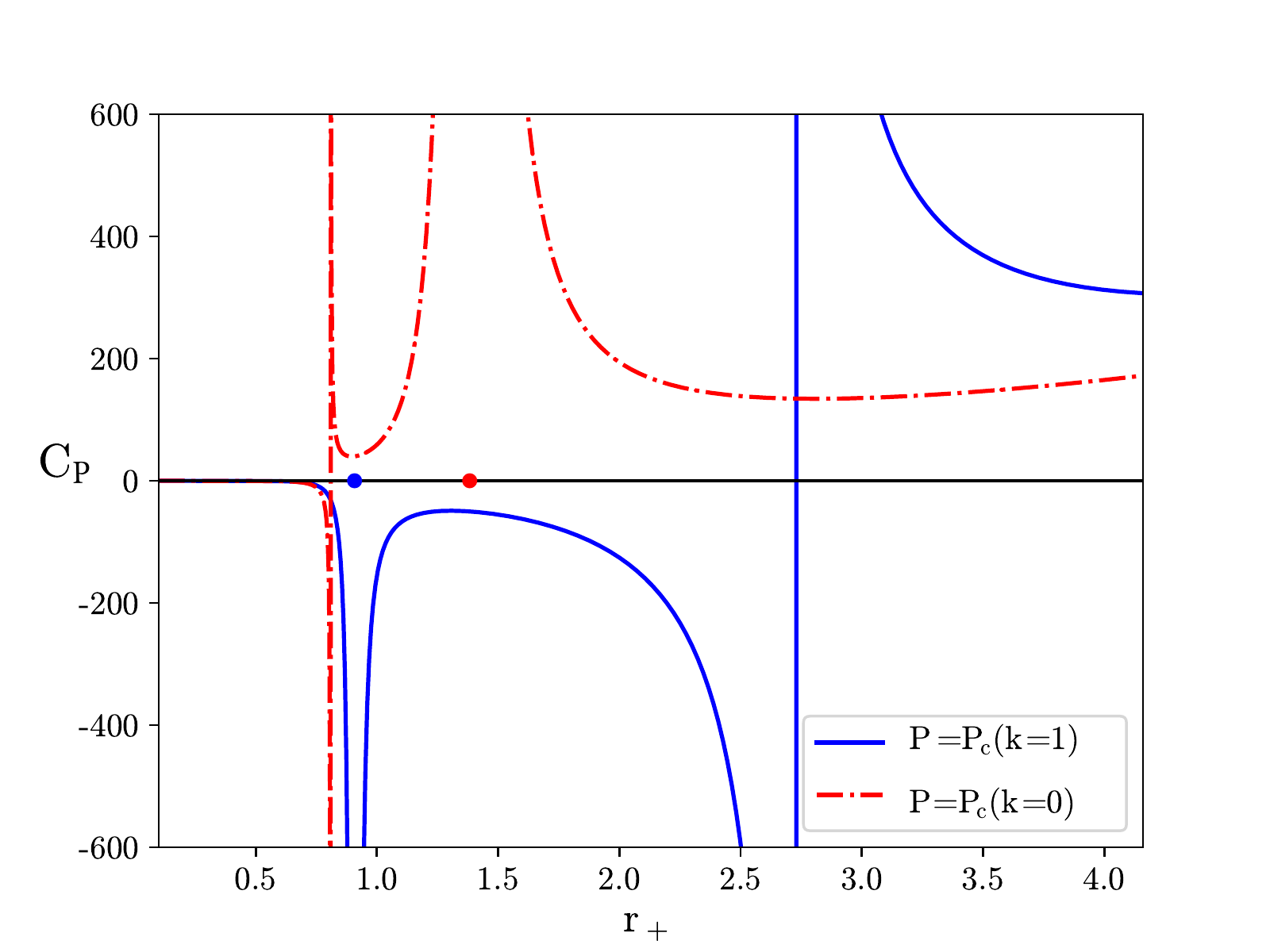}
  \caption {$C_{\rm P}$ as a function of $r_{+}$ for the EHAdS BH. The red curves are for $k = 0$, the blue curves are for $k = 1$. The blue point and red point represent first critical point and second critical point, respectively. We take EH parameter $a=1$ and BH charge $Q=0.6$.}\label{fig:3}
\end{figure}

The heat capacity of the EHAdS BH is derived from
\begin{eqnarray}
C_{\rm P}&=&T\big(\frac{\partial S}{\partial T}\big)_{P}, \\
&=&\frac{2\pi r_{+}^2(aQ^4-4Q^2r_{+}^4+4r_{+}^6+32P\pi r_{+}^8)}{-7aQ^4+12Q^2r_{+}^4-4r_{+}^6+32P\pi r_{+}^8}.\nonumber
\label{2-0-11}
\end{eqnarray}
Figure \ref{fig:3} illustrates $C_{\rm P}$ as a function of $r_{+}$. We can observe that $C_{\rm P}$ curves are discontinuity. A sign change for both critical points in the $C_{\rm P}$ diagram, showing the thermodynamic instability of the EHAdS BH.

\par
The $P-V$ and $T-S$ type Maxwell's equal area law can be given by \cite{15}
\begin{equation}
P_{\rm i}(V_2-V_1)=\int_{V_1}^{V_2}P{\rm d}V,
\label{2-0-12}
\end{equation}
\begin{equation}
T_{\rm i}(S_2-S_1)=\int_{S_1}^{S_2}T{\rm d}S,
\label{2-0-13}
\end{equation}
where $P_{\rm i}$ (or $T_{\rm i}$) represents the pressure (or temperature) of isobar (or isotherm), $V$ is the thermodynamic volume, $S$ is the entropy, the subscript 1 (or 2) stands for the start (or end) phase of isobaric (or isothermal) process. According to Eqs. (\ref{2-0-5}) and (\ref{2-0-12}), we have
\begin{eqnarray}
P_{\rm i}&=&\frac{T_{\rm pt}}{2r_{\rm 1}}-\frac{1}{8\pi r_{\rm 1}^2}+\frac{Q^2}{8\pi r_{\rm 1}^4}-\frac{aQ^4}{32\pi r_{\rm 1}^8},
\label{2-0-14}\\
P_{\rm i}&=&\frac{T_{\rm pt}}{2r_{\rm 2}}-\frac{1}{8\pi r_{\rm 2}^2}+\frac{Q^2}{8\pi r_{\rm 2}^4}-\frac{aQ^4}{32\pi r_{\rm 2}^8},
\label{2-0-15}\\
2P_{\rm i}&=&\frac{3\Big(r_{\rm 2}^2x\big(2\pi r_{\rm 2}T_{\rm pt}(1+x)-1\big)+Q^2\Big)}{4\pi r_{\rm 2}^4x(1+x+x^2)}\nonumber \\
&-&\frac{3aQ^4(1+x+x^2+x^3+x^4)}{80\pi r_{\rm 2}^8x^5(1+x+x^2)}.
\label{2-0-16}
\end{eqnarray}
Using the Eqs. (\ref{2-0-14})-(\ref{2-0-16}), we have,
\begin{eqnarray}
r_{\rm 2}^2=\frac{1+4x+x^2}{6x^2}A,
\label{2-0-19}
\end{eqnarray}
where $x \equiv r_{1}/r_{2}$ ($0<x<1$), $A \equiv 2Q^2[1+2\cos(\arccos(1-(27aB)/(40Q^2))/3-2k\pi/3)]$ and $B \equiv (5+20x+29x^2+32x^3+29x^4+20x^5+5x^6)/(1+4x+x^2)^3$.
The phase transition temperature $T_{\rm pt}$ can be derived as
\begin{eqnarray}
T_{\rm pt}&=&\frac{(1+x)\big(r_{\rm 2}^2x^2-Q^2(1+x^2)\big)}{4\pi r_{\rm 2}^3x^3}\nonumber \\
&+&\frac{(1+x)aQ^4(1+x)(1+x^2)(1+x^4)}{16\pi r_{\rm 2}^7x^7}.
\label{2-0-20}
\end{eqnarray}
We also construct the $T-S$ type equal area law by utilizing Eqs. (\ref{2-0-4}) and (\ref{2-0-13}). The phase transition pressure $P_{\rm pt}$ can be written as
\begin{eqnarray}
P_{\rm pt}&=&\frac{r_{\rm 2}^2x^2-Q^2(1+x+x^2)}{8\pi r_{\rm 2}^4x^3}\nonumber \\
&+&\frac{aQ^4(1+x+x^2+x^3+x^4+x^5+x^6)}{32\pi r_{\rm 2}^8x^7}.
\label{2-0-21}
\end{eqnarray}
We define a parameter $\chi_{\rm Z} \equiv \frac{{Z-Z_{c1}}}{{Z_{\rm c0}-Z_{\rm c1}}}$ ($0<\chi_{\rm Z}<1$) to measure the temperature (pressure) of different phase transition branches, where $Z$ is the phase transition temperature $T_{\rm pt}$ (pressure $P_{\rm pt}$). The subscripts $c0$ and $c1$ denote the $k$ value of two critical points.

\par
\begin{figure*}[htbp]
  \centering
  \includegraphics[width=7.5cm,height=5cm]{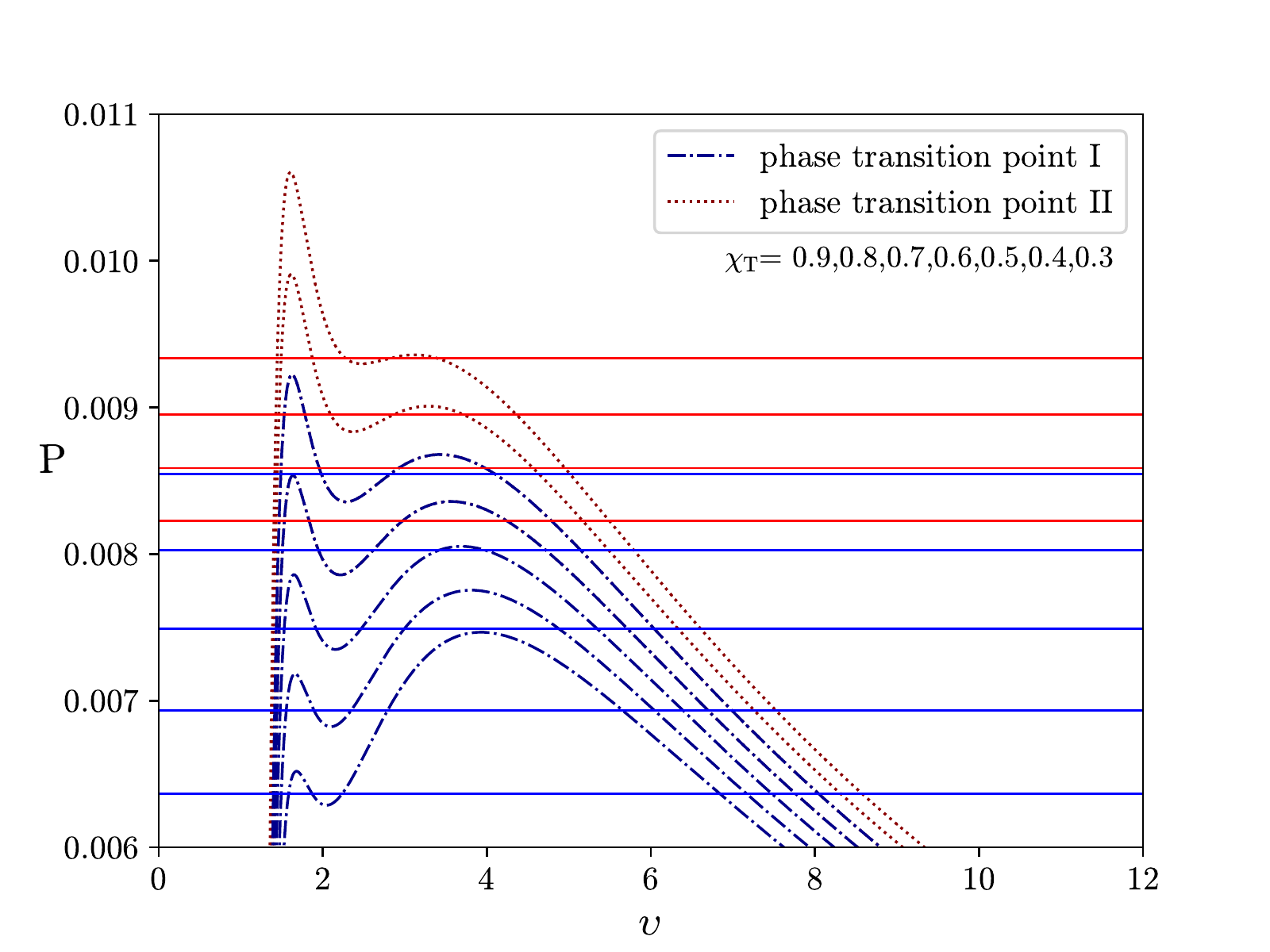}
  \hspace{0.5cm}
  \includegraphics[width=7.5cm,height=5cm]{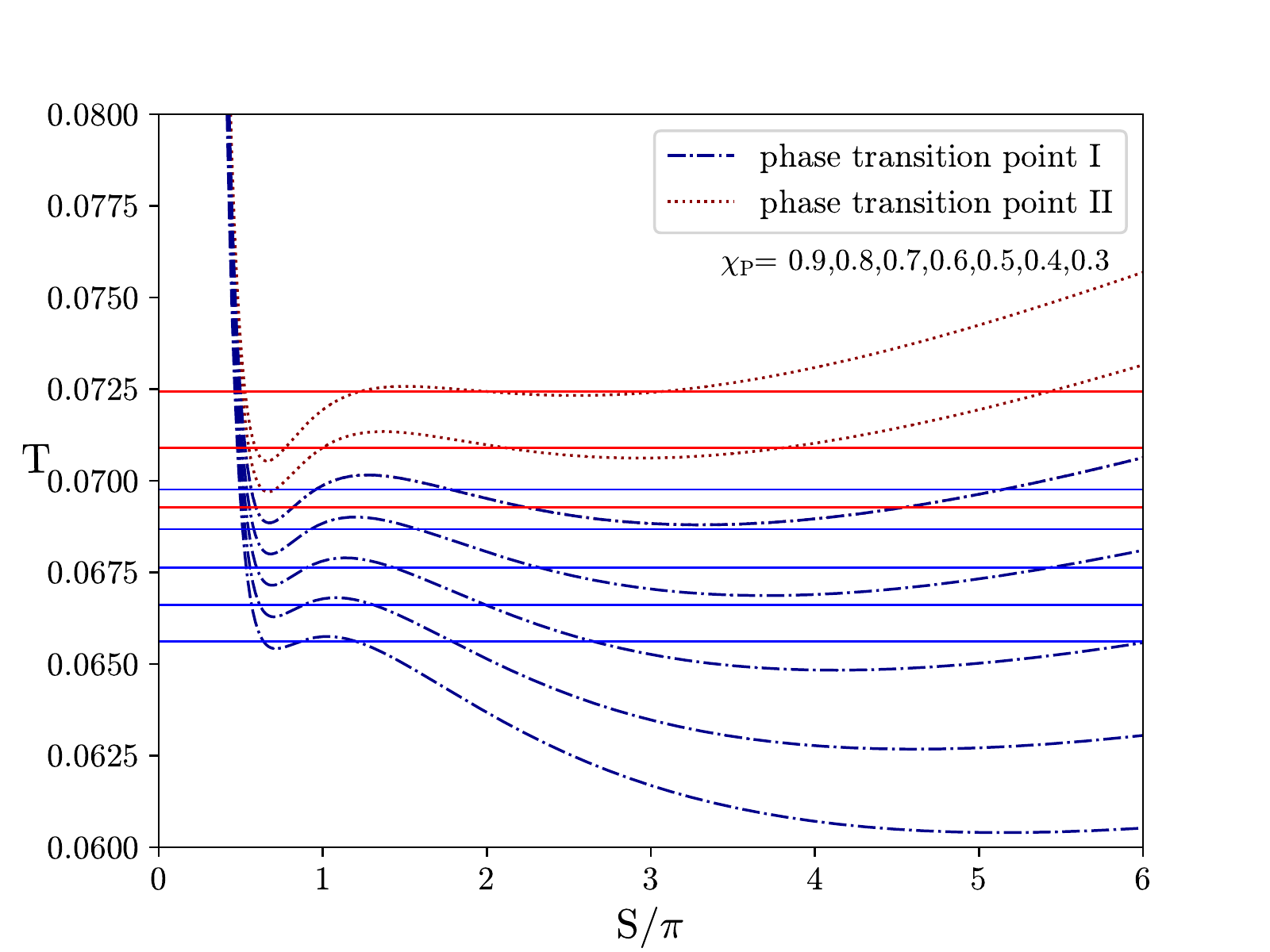}
  \caption {The Maxwell¡¯s equal area law of the EHAdS BH. The blue and red solid lines present the first and second phase transition branches. We take EH parameter $a=1$ and BH charge $Q=0.6$.}\label{fig:4}
\end{figure*}
\begin{figure}[htbp]
  \centering
  \includegraphics[width=7.5cm,height=5cm]{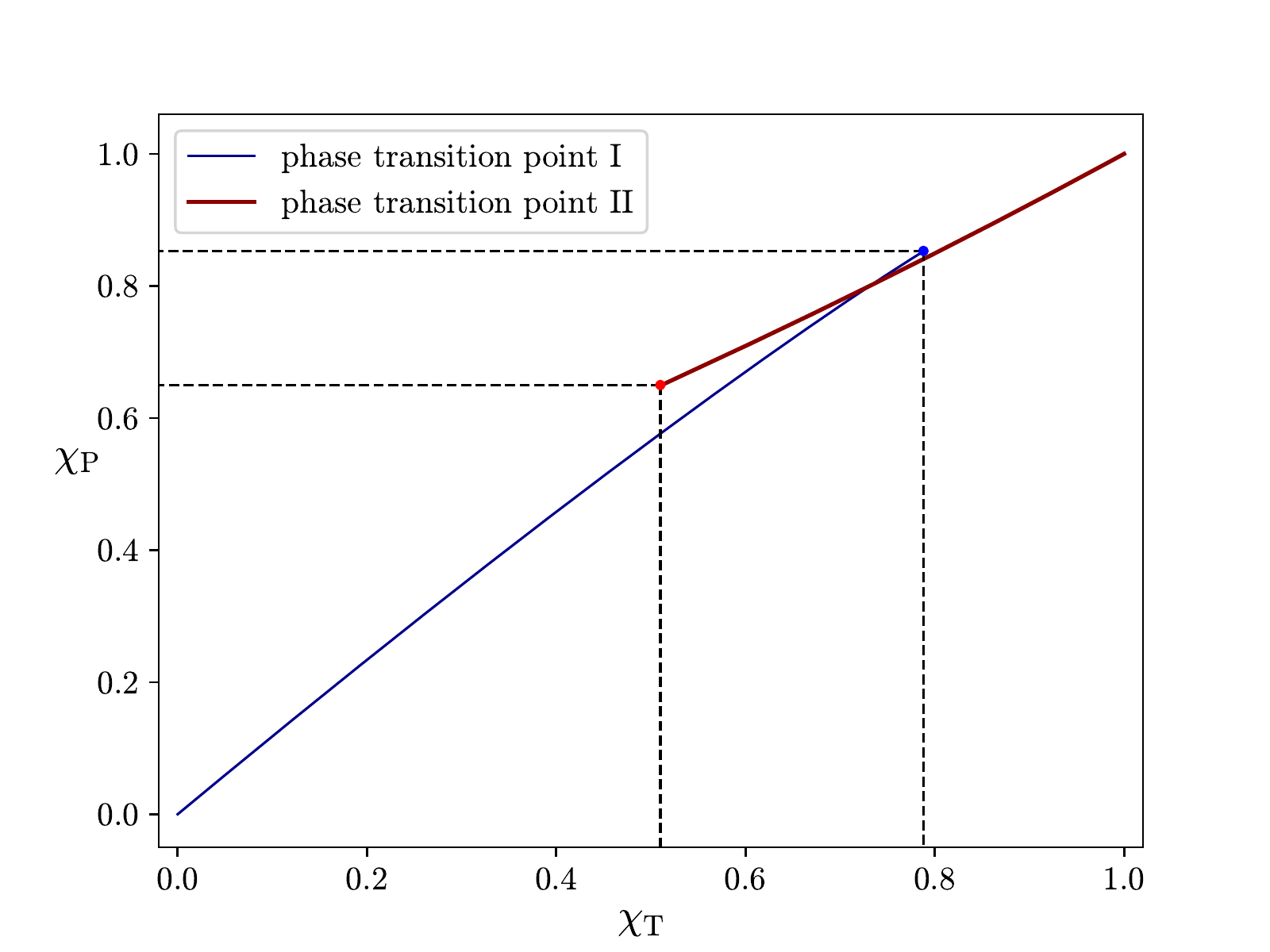}
  \caption {The blue and red line present the first and second phase transition branches for the EHAdS BH. We take EH parameter $a=1$ and BH charge $Q=0.6$.}\label{fig:5}
\end{figure}

Using Eqs. (\ref{2-0-19})-(\ref{2-0-21}), the isobaric (isothermal) curves of the EHAdS BH on the $P-V$ ($T-S$) plane are shown in Fig.\ref{fig:4}. The length of the isothermal (isobaric) horizontal segment increases gradually with the temperature (pressure) increase for the first phase transition branch, while it is different from the second phase transition branch. Furthermore, instability may occur before (the red dash curves bulge) or after (the blue dash curves bulge) phase transition. Figure \ref{fig:5} shows reentry the phase transition region. Only one stable phase transition is remained. The $\chi_{\rm P}$ and $\chi_{\rm T}$ are in the ranges of (0.65,0.85) and (0.51,0.79), respectively.

\section{The EHAdS BH with high-order QED correction and thermodynamic characteristics}
\label{sec:3}
\par
\subsection{The solution of the corrected EHAdS BH}
\label{sec:3-1}
\par
Classical electrodynamics is modified for strong electromagnetic fields because of the self-interaction of photons \cite{24}. QED is considered to characterize the nonlinear due to loop corrections and vacuum birefringence's effect. The main feature of the QED correction ensures that particles' electric field and electrostatic energy comply with Coulomb's law when $r\rightarrow\infty$ \cite{21}. Here, we consider the QED high-order correction. This would relax the limit of the energy-momentum tensor and the NEM field limit of the EHAdS BH. Additionally, the high-order QED correction may also affect the phase transition of the EHAdS BH.
\par
The action of the EHAdS BH is given by \cite{22}
\begin{equation}
I=\int {\rm d}^4\sqrt{-g}\Big(\frac{1}{2\kappa^2}(R-2\Lambda)-\mathcal{L(\mathcal F,\mathcal G)}\Big),
\label{3-0-1}
\end{equation}
where $\kappa^2 \equiv 8\pi G$, R is the Ricci scalar. The one-loop effective Lagrangian density of the EH nonlinear electrodynamics $\mathcal{L(\mathcal F,\mathcal G)} = -\mathcal F+\frac{a}{2}\mathcal F^2+\frac{7a}{8}\mathcal G^2 $, and $\mathcal F =\frac{1}{4}F_{\rm \mu\nu}F^{\rm \mu\nu}$, $\mathcal G= \frac{1}{4}F_{\rm \mu\nu}{\widetilde{F}}^{\rm \mu\nu}$, where $F_{\rm \mu\nu}$ is the electromagnetic field strength tensor and its dual ${\widetilde{F}}^{\rm \mu\nu} = \frac{1}{2\sqrt{-g}}\epsilon_{\rm \mu\nu\sigma\rho}F^{\rm \sigma\rho}$. The field equations read as \cite{22}
\begin{eqnarray}
\nabla_{\mu}P^{\rm \mu\nu}=0,~~~~G_{\rm \mu\nu}+\Lambda g_{\rm \mu\nu}=\kappa^2T_{\rm \mu\nu},
\label{3-0-ee}
\end{eqnarray}
where $P^{\rm \mu\nu} = (1-a\mathcal F)F_{\rm \mu\nu}-{\widetilde{F}}^{\rm \mu\nu}\frac{7a}{4}\mathcal G$. Similarly, considering only the electric field $(B=0)$, the above equations can be rewritten as
\begin{eqnarray}
{\rm\partial_{\rm r}}\Bigg(r^2E\Big(\frac{aE^2}{2}+1\Big)\Bigg)=0,
\label{3-0-2}\\
G_{\rm \mu\nu}+\Lambda g_{\rm \mu\nu}=\kappa^2T_{\rm \mu\nu}.
\label{3-0-3}
\end{eqnarray}
Assuming that the EH nonlinear electric field near the BH is generated by the EHAdS BH charge, the Eq. (\ref{3-0-2}) is replaced with the active form of the EH nonlinear electric field, i.e. ${\rm\partial_{\rm r}}\Big(r^2E(\frac{aE^2}{2}+1)\Big)=4\pi\rho$, where $\rho$ is charge density of the EHAdS BH. This effect can also be achieved by imposing Coulomb's law constraints on the EH nonlinear electric field $\frac{\partial\mathcal L}{\partial\boldsymbol E} = \frac{Q}{r^3}\boldsymbol r$ \cite{25}. Hence, $E(r)$ is \cite{21}
\begin{equation}
E(r)=\frac{\sqrt{8}}{\sqrt{3a}}sinh\Bigg(\frac{1}{3}ln\Bigg(\frac{\sqrt{27a}Q}{\sqrt{8}r^2}+\sqrt{\frac{27aQ^2}{8r^4}+1}\Bigg)\Bigg).
\label{3-0-4}
\end{equation}
Utilizing the condition $r\rightarrow\infty$, $E(r)$ expansion form is
\begin{equation}
E(r)=\frac{Q}{r^2}-\frac{aQ^3}{2r^6}+\frac{3a^2Q^5}{4r^{10}}-\frac{12a^3Q^7}{8r^{14}}+\mathcal{O}(r^{-18}).
\label{3-0-5}
\end{equation}
The $(0,0)$ component of the energy-momentum tensor $T_{\mu\nu}$ is
\begin{eqnarray}
T_{\rm 00}&&=\frac{E^2}{2}\Big(1+\frac{3aE^2}{4}\Big)\nonumber\\
&&=\frac{Q^2}{2r^4}-\frac{aQ^4}{8r^8}+\frac{a^2Q^6}{8r^{12}}-\frac{3a^3Q^8}{16r^{16}}+\mathcal{O}(r^{-20}).
\label{3-0-6}
\end{eqnarray}
Considering only the first two terms of Eq. (\ref{3-0-6}) and basing on Eq. (\ref{3-0-2}), we have
\begin{equation}
\frac{{\rm d}m}{{\rm d}r}=\frac{Q^2}{2r^2}-\frac{aQ^4}{8r^6}+\frac{\Lambda r^2}{2}.
\label{3-0-7}
\end{equation}
The metric potential (Eq. \ref{2-0-2}) of the EHAdS BH is obtained by integrating the above equation. Treating the third-order and subsequent terms as the high-order QED correction, Eq. (\ref{3-0-6}) can be written as
\begin{equation}
T_{00}=\frac{Q^2}{2r^4}-\frac{aQ^4}{8r^8}+\frac{\beta a^2Q^6}{8r^{12}},
\label{3-0-8}
\end{equation}
where the 3th term is the higher-order terms and $\beta\sim 1$. Using the same method, Eq. (\ref{3-0-7}) can be rewritten as
\begin{eqnarray}
\frac{{\rm d}m'}{{\rm d}r}=\frac{{\rm d}m}{{\rm d}r}+\frac{\beta a^2Q^6}{8r^{10}}.
\label{3-0-9}
\end{eqnarray}
Hence, the metric potential $f'(r)$ of the corrected EHAdS BH is
\begin{eqnarray}
f'(r)=f(r)+\frac{\beta a^2Q^6}{36r^{10}}.
\label{3-0-10}
\end{eqnarray}

\subsection{Thermodynamics of the EHAdS BH within the high-order QED correction framework}
\label{sec:3-2}
\par
\begin{figure}[htbp]
  \centering
  \includegraphics[width=7.5cm,height=5cm]{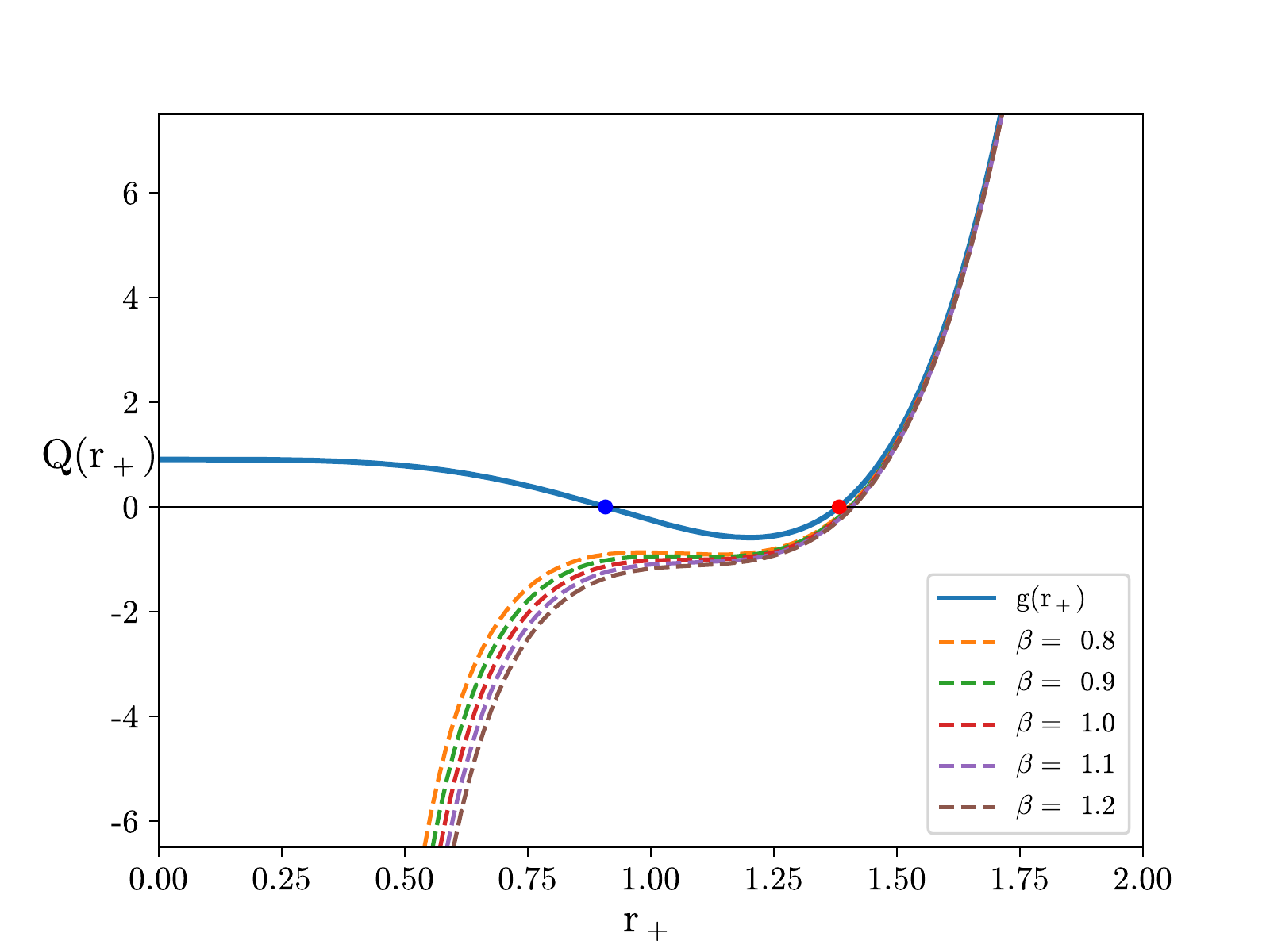}
  \caption {$Q({r_{+}})$ as a function of $r_{+}$ under different values of parameter $\beta$ for the EH parameter $a=1$ and BH charge $Q=0.6$.}\label{fig:6}
\end{figure}

In this scenario, the corrected mass $M'$ is
\begin{eqnarray}
M'=M+\frac{\beta a^2Q^6}{72r_{+}^9}.
\label{3-0-11}
\end{eqnarray}
The corrected Hawking temperature and equation of the state can be written as
\begin{eqnarray}
T'&=&T-\frac{\beta a^2Q^6}{16\pi r_{+}^{11}},
\label{3-0-12}\\
P'&=&P+\frac{\beta a^2Q^6}{32\pi r_{+}^{12}}.
\label{3-0-13}
\end{eqnarray}
Utilizing Eq. (\ref{3-0-13}) and critical conditions, one can get
\begin{eqnarray}
Q(r_{+})\equiv g(r_{+})-\frac{33a^2Q^6\beta}{2r_{+}^4}=0.
\label{3-0-14}
\end{eqnarray}
$Q(r_{+})$ as a function of $r_{+}$ is plotted in Figure \ref{fig:6}. $Q(r_{+})$ function curve intersects the horizontal axis once for different $\beta$ values, indicating that there is one phase transition of the corrected EHAdS BH.

Similarly, the critical thermodynamic quantities of the corrected EHAdS BH are obtained, i.e.
\begin{eqnarray}
{T'_{\rm c}}&=&-\frac{3a^2Q^6\beta}{4\pi {r'_{\rm c}}^{11}}+\frac{aQ^4}{2\pi {r'_{\rm c}}^7}-\frac{Q^2}{\pi {r'_{\rm c}}^3}+\frac{1}{2\pi {r'_{\rm c}}},
\label{3-0-15}\\
{P'_{\rm c}}&=&-\frac{11a^2Q^6\beta}{32\pi {r'_{\rm c}}^{12}}+\frac{7aQ^4}{32\pi {r'_{\rm c}}^8}-\frac{3Q^2}{8\pi {r'_{\rm c}}^4}+\frac{1}{8\pi {r'_{\rm c}}^2},
\label{3-0-16}\\
{r'_{\rm c}}^2&=&r_{\rm c}^2+\frac{-C+(-1)^k\sqrt{33a^2Q^6D\beta+C^2}}{D},
\label{3-0-17}
\end{eqnarray}
where $C \equiv 14aQ^4r_{\rm c}^2-24Q^2r_{\rm c}^6+5r_{\rm c}^8$, and $D \equiv 14aQ^4-72Q^2r_{\rm c}^4+20r_{\rm c}^6$. If $\beta \rightarrow 0$, Eqs. (\ref{3-0-15})-(\ref{3-0-17}) degenerate into the EHAdS BH case (Eqs. \ref{2-0-6}-\ref{2-0-8}). In case of $\beta \rightarrow 1$, the $k$ value should be zero to ensure ${r'_{\rm c}}^2$ is positive. The critical radius can be approximated by the limit of $a \rightarrow 0$,
\begin{eqnarray}
&\lim_{a\to0}{r'_{\rm c}}^2&=\lim_{a\to0}r_{\rm c}^2\nonumber\\
&=4Q^2\cos\Big(&\frac{1}{3}\arccos\big(1-\frac{7a}{16Q^2}\big)\Big)+2Q^2=6Q^2.
\label{3-0-aa}
\end{eqnarray}
In this limit, we have
\begin{eqnarray}
&{\upsilon '_{\rm c}}&\approx 2\sqrt{6}Q,\\
\label{3-0-bb}
&{T'_{\rm c}}&\approx\frac{1}{3\sqrt{6}\pi Q}+\frac{a}{432\sqrt{6}\pi Q^3}-\frac{a^2\beta}{10368\sqrt{6}\pi Q^5},\\
\label{3-0-bb-1}
&{P'_{\rm c}}&\approx\frac{1}{96\pi Q^2}+\frac{7a}{41472\pi Q^4}-\frac{11a^2\beta}{1492992\pi Q^6}.
\label{3-0-bb-2}
\end{eqnarray}
Therefore, the universal constant $\varepsilon'$ of the corrected EHAdS BH at $a\to0$ is
\begin{eqnarray}
\varepsilon'\approx\frac{3}{8}a^0+\frac{1}{288Q^2}a^1-\frac{(2+13\beta)}{82944Q^4}a^2+\mathcal{O}(a^3).
\label{3-0-cc}
\end{eqnarray}
One can see that the $\varepsilon' = 3/8$ for the zeroth-order $a$ term, which is same as the vdW system, the 1st-order term is $1/288$, and the 2nd-order term is 13/82994 in comparison that without the high-order QED correction. For simplicity, we have omitted the 3rd-order and subsequent correction.

\par
Furthermore, we calculate the critical exponents in the correction situation. In the reduced parameter space, the reduced equation of the state can be given as
\begin{eqnarray}
p'&&=\frac{\tau'}{\varepsilon'\nu'}+\frac{1}{\pi P'_{\rm c} {\upsilon'_{\rm c}}^2}\Big(-\frac{1}{2\nu'^2}+\frac{1}{12\nu'^4}\Big)\nonumber\\
&&-\frac{1}{\pi P'_{\rm c} {\upsilon'_{\rm c}}^2}\Big(\frac{a}{1728 Q^2 \nu'^8}-\frac{\beta a^2 }{62208 Q^4 \nu'^{12}}\Big),
\label{A-1}
\end{eqnarray}
where
\begin{eqnarray}
p'=\frac{P'}{P'_{\rm c}}, ~~~\tau'=\frac{T'}{T'_{\rm c}}, ~~~\nu'=\frac{\upsilon'}{\upsilon'_{\rm c}}.
\label{A-2}
\end{eqnarray}
Utilizing Eqs. (\ref{3-0-bb})-(\ref{3-0-bb-2}), the Eq. (\ref{A-1}) can be rewritten as
\begin{eqnarray}
p'=\frac{\tau'}{\varepsilon'\nu'} - \frac{2}{\nu'^2}+\frac{1}{3\nu'^4}+h(\nu'),
\label{A-3}
\end{eqnarray}
where $h(\nu')$ is
\begin{eqnarray}
h(\nu')&=&\frac{a(432Q^2-7a)(42\nu'^{6}-7\nu'^{4}-3)}{559872\nu'^{12}Q^4}\nonumber\\
&-&\frac{\beta a^2(66\nu'^{10}-11\nu'^{8}-3)}{46656\nu'^{12}Q^4}+\mathcal{O}[a^3].
\label{A-4}
\end{eqnarray}
By introducing reduced parameters $t'$ and $\omega'$, which are defined as
\begin{equation}
{t'}\equiv{\tau'}-1,\qquad {\omega'}\equiv({\nu'}-1)^{1/3},
\label{A-5}
\end{equation}
one can get the expansion of the reduced state equation near the critical point ($t'\to0$, $\omega'\to0$),
\begin{eqnarray}
p'(t',\omega')&&\approx 1+\frac{1}{\varepsilon'}t'-\frac{1}{\varepsilon'}\omega't'\nonumber\\
&&+\Big(\frac{4}{3}-\frac{1}{\varepsilon'}+\frac{{h'}^{(3)}(1)}{6}\Big){\omega'}^3+\mathcal{O}({\omega'}^2t',{\omega'}^4),
\label{A-6}
\end{eqnarray}
where $h^{(3)}(1)$ is given as
\begin{eqnarray}
{h'}^{(3)}(1)=\frac{83a(732Q^2-7a)}{139968Q^4}-\frac{\beta a^2 131}{5832Q^4}.
\label{A-7}
\end{eqnarray}
With the equal area law, the volumes of small BH ($\omega'_{\rm 1}$) and large BH ($\omega'_{\rm 2}$) satisfy
\begin{equation}
\label{A-8}
\int_{\omega'_{\rm 1}}^{\omega'_{\rm 2}}\omega'{\rm d}p'=0,
\end{equation}
where ${\rm d}p' = \Big(-\frac{t'}{\varepsilon'}+(4-\frac{3}{\varepsilon'}+\frac{{h'}^{(3)}(1)}{2}){\omega'}^2\Big){\rm d}\omega'$. Thus, the above equation has a unique nontrivial solution, that is
\begin{equation}
\omega'_{\rm 1}=-\omega'_{\rm 2}=\sqrt{\Big(-\frac{4}{3\varepsilon'}+\frac{1}{{\varepsilon'}^2}-\frac{{h'}^{(3)}(1)}{6\varepsilon'}\Big)\big(-t'\big)}.
\label{A-9}
\end{equation}
We known that the critical exponents of the AdS BH can be given by \cite{4}
\begin{eqnarray}
C_{\rm v}&=&T\Big(\frac{\partial S}{\partial T}\Big)_{\rm v}\propto|t|^{-\mathcal{\alpha}},~~~~\eta=V_{2}-V_{1}\propto|-t|^{\mathcal{\beta}},\\
\label{A-10}
\kappa_{T}&=&-\frac{1}{V}\Big(\frac{\partial V}{\partial P}\Big)\propto|t|^{-\mathcal{\gamma}},~~~P-P_{c}\propto|V-V_{c}|^{\mathcal{\delta}}.
\label{A-10-1}
\end{eqnarray}
For the corrected case, the heat capacity at constant volume is $C_{\rm v'}=0$, inferring that the first critical exponent as $\alpha=0$. According to Eq. (\ref{A-9}), we have
\begin{eqnarray}
\eta'=\upsilon'_{\rm c}(\omega'_{\rm 1}-\omega'_{\rm 2})=2\upsilon'_{\rm c}\omega'_{\rm 1}\propto|-t'|^{1/2}.
\label{A-12}
\end{eqnarray}
The second critical exponent is $\beta=1/2$. The third and fourth critical exponents are $\gamma=1$ and $\delta=3$ since
\begin{eqnarray}
\kappa_{\rm T'}&=&-\frac{1}{V}\Big(\frac{\partial V}{\partial P}\Big)=-\frac{1}{P'_{\rm c}(\omega'+1)}\frac{{\rm d}\omega'}{{\rm d}p'}\propto\frac{\varepsilon'}{t'},\\
\label{A-13}
|p'&-&1|=\Big(\frac{4}{3}-\frac{1}{\varepsilon'}+\frac{{h'}^{(3)}(1)}{6}\Big){\omega'}^3 \propto|\omega'|^3.
\label{A-14}
\end{eqnarray}
The critical exponents $(\alpha,~\beta,~\gamma,~\delta)$ are $(0,~1/2,~1,~3)$, which are equivalent to the critical exponents of the vdW system \cite{4}.

\par
The heat capacity is
\begin{eqnarray}
&&C_{\rm P'}=T'\big(\frac{\partial S'}{\partial T'}\big)_{P'}\\
&&=\frac{2\pi r_{+}^2\big(aQ^4r_{+}^4+4r_{+}^8(-Q^2+r_{+}^2+8P'\pi r_{+}^4)-a^2Q^6\beta\big)}{-7aQ^4r_{+}^4+4r_{+}^8(3Q^2-r_{+}^2+8P'\pi r_{+}^4)+11a^2Q^6\beta}.\nonumber
\label{3-0-dd}
\end{eqnarray}
\begin{figure}[htbp]
  \centering
  \includegraphics[width=7.5cm,height=5cm]{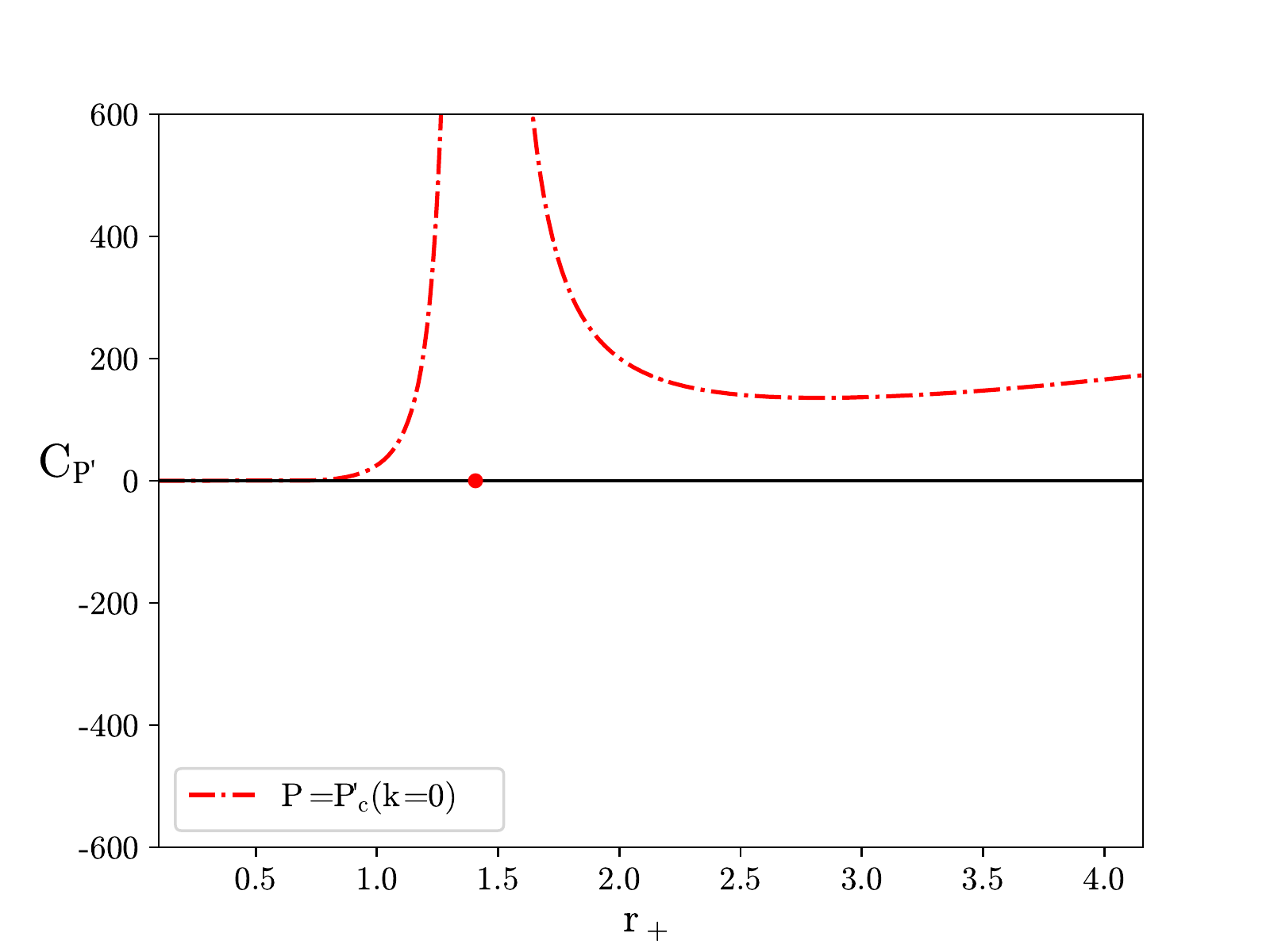}
  \caption {$C_{\rm P'}$ as a function of $r_{+}$ of the corrected EHAdS BH. The red point is the critical point. We take EH parameter $a=1$, BH charge $Q=0.6$ and parameter $\beta=1$.}\label{fig:7}
\end{figure}
Figure \ref{fig:7} illustrates the $C_{\rm P'}$ as a function of $r_{+}$. It is found that the sign of the $C_{\rm P'}$ does not change, indicating that the thermodynamic instability of the BH disappears. According to Eqs. (\ref{2-0-12}) and (\ref{3-0-13}), the $P-V$ type Maxwell¡¯s equal area law in this scenario is constructed. We have
\begin{eqnarray}
P'_{\rm i}&=&P_{\rm i}(r'_{\rm 1})+\frac{\beta a^2Q^6}{32\pi {r'_{\rm 1}}^{12}},
\label{3-0-18}\\
P'_{\rm i}&=&P_{\rm i}(r'_{\rm 2})+\frac{\beta a^2Q^6}{32\pi {r'_{\rm 2}}^{12}},
\label{3-0-19}\\
2P'_{\rm i}&=&2P_{\rm i}(r'_{\rm 2},x')+\frac{\beta a^2Q^6(\sum_{i=0}^{8}{x'}^{i})}{48\pi {r'_{\rm 2}}^{12}x'^9(1+x'+x'^2)},
\label{3-0-20}
\end{eqnarray}
where $x' \equiv r'_{\rm 1}/r'_{\rm 2}~(0<x'<1)$, and $P_{\rm i}(r'_{\rm 1})$, $P_{\rm i}(r'_{\rm 2})$ , $2P_{\rm i}(r'_{\rm 2},x')$ are from Eqs. (\ref{2-0-14})-(\ref{2-0-16}) by replacing $r_{\rm 1/2}$ with $r'_{\rm 1/2}$, $x$ with $x'$, and $T_{\rm pt}$ with $T'_{\rm pt}$. Based on Eqs. (\ref{3-0-18}) and (\ref{3-0-20}), one can get

\begin{small}
\begin{eqnarray}
\label{3-0-23}
&{r'_{\rm 2}}^2&=r_{\rm 2}^{2}+\frac{-2Er_{\rm 2}^{2}-4Fr_{\rm 2}^{6}-5r_{\rm 2}^{8}}{2(E+6Fr_{\rm 2}^{4}+10r_{\rm 2}^{6})}\\
&+&(-1)^k\frac{\sqrt{\Big(2Er_{\rm 2}^{2}+4Fr_{\rm 2}^{6}+5r_{\rm 2}^{8}\Big)^2-4G(E+6Fr_{\rm 2}^{4}+10r_{\rm 2}^{6})\beta}}{2E+12Fr_{\rm 2}^{4}+20r_{\rm 2}^{6}},\nonumber
\end{eqnarray}
\end{small}
in which $E \equiv aQ^4(5+20x'+29x'^2+32x'^3+29x'^4+20x'^5+x'^6)/(20x'^6)$, $F \equiv -Q^2(1+4x'+x'^2)/x'^2$, and $G \equiv -a^2Q^6(3+12x'+19x'^2+24x'^3+27x'^4+28x'^5+27x'^6+24x'^7+19x'^8+12x'^9+3x'^{10})/(12x'^{10})$. The phase transition temperature $T'_{\rm pt}$ is
\begin{eqnarray}
\label{3-0-21}
T'_{\rm pt}=T_{\rm pt}(r'_{\rm 2},x')-\frac{\beta a^2Q^6(\sum_{i=0}^{11}x'^{i})}{16\pi {r'_{\rm 2}}^{11}x'^{11}}.
\end{eqnarray}
The $T-S$ type equal area law is derived from Eqs. (\ref{2-0-13}) and (\ref{3-0-12}). The phase transition temperature $P'_{\rm pt}$ is
\begin{eqnarray}
P'_{\rm pt}=P_{\rm pt}(r'_{\rm 2},x')-\frac{\beta a^2Q^6(\sum_{i=0}^{10}x'^{i})}{32\pi {r'_{\rm 2}}^{12}x'^{11}}.
\label{3-0-24}
\end{eqnarray}
$T_{\rm pt}(r'_{\rm 2},x')$ and $P_{\rm pt}(r'_{\rm 2},x')$ are from the Eqs. (\ref{2-0-20}) and (\ref{2-0-21}) by replacing $r_{\rm 2}$ with $r'_{\rm 2}$ and $x$ with $x'$, respectively.

\begin{figure*}[htbp]
  \centering
  \includegraphics[width=7.5cm,height=5cm]{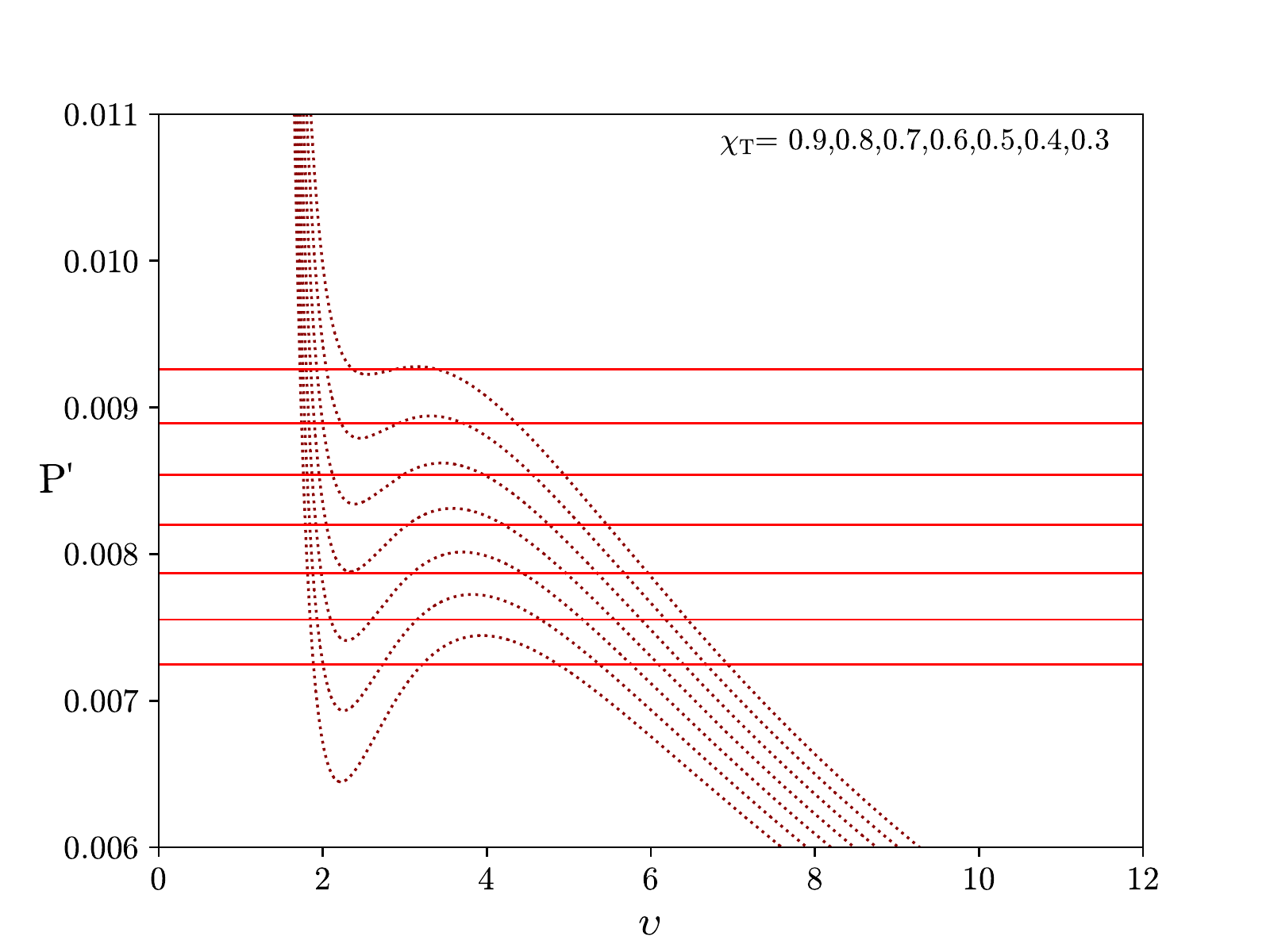}
  \hspace{0.1cm}
  \includegraphics[width=7.5cm,height=5cm]{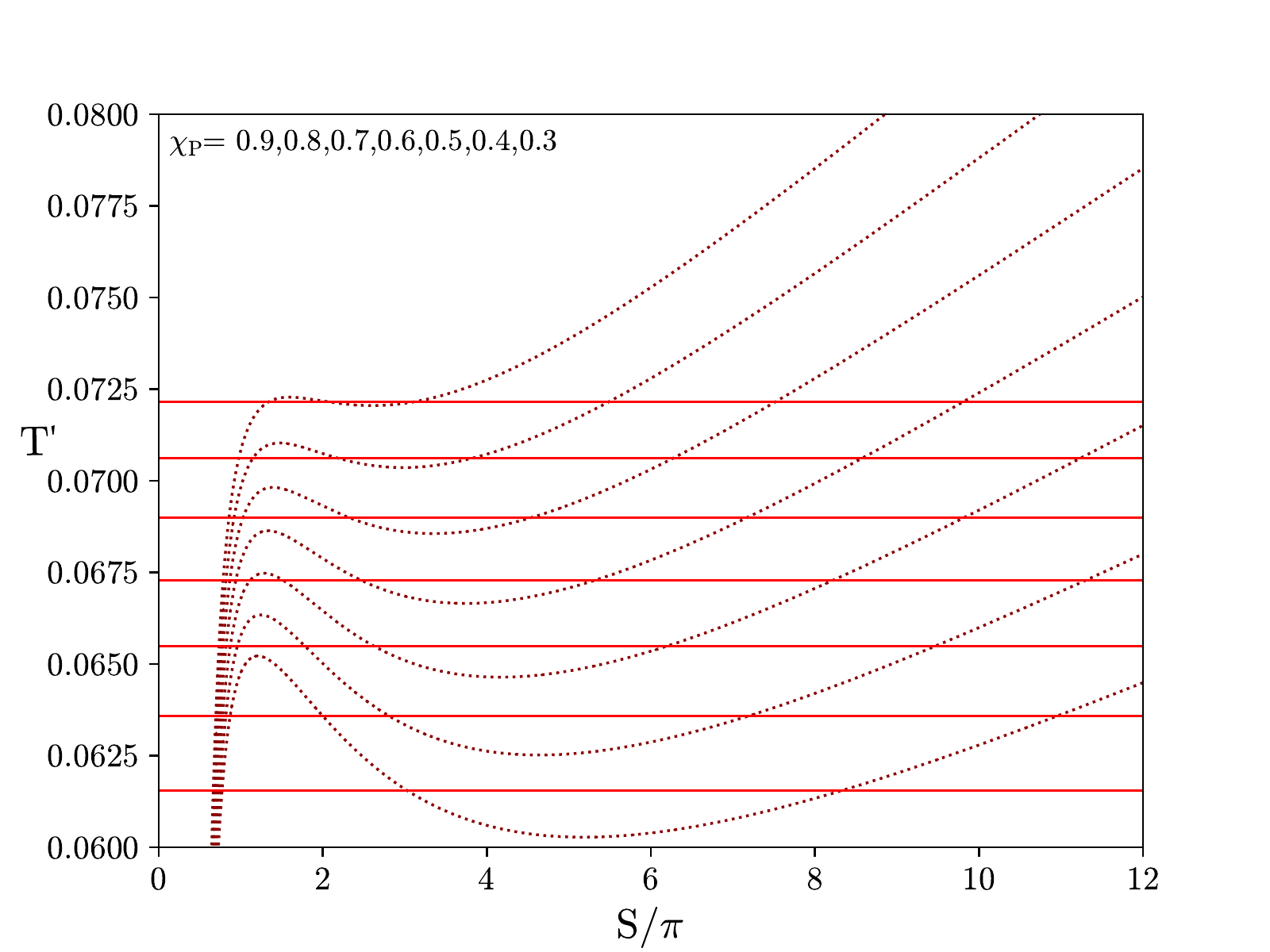}
  \caption {The Maxwell¡¯s equal area law of the corrected EHAdS BH. We take EH parameter $a=1$, BH charge $Q=0.6$ and parameter $\beta=1$.}\label{fig:8}
\end{figure*}

Using Eqs. (\ref{3-0-23}), (\ref{3-0-21}), and (\ref{3-0-24}), the isobaric (and isothermal) curves of the corrected EHAdS BH on the $P-V$ and $T-S$ planes are shown in Fig.\ref{fig:8}. One can observe that the length of the isothermal (or isobaric) horizontal segment decreases gradually with the increase in temperature (or pressure). It is the same as the vdW system and the charged AdS BH, while the instability feature of the EHAdS BH disappears.

\par
On the other hand, the Ruppeiner geometry of the BH system can reveal the microstructure of BH phase transition. We analyze the microscopic phase transition behavior of the EHAdS BHs by investigating the Ruppeiner geometry. The normalized scalar curvature $R_{\rm N}$ is given by \cite{9}
\begin{equation}
R_{\rm N}=\frac{(\partial_{\rm V}P)^2-T^2(\partial_{\rm V,T})^2+2T^2(\partial_{\rm V}P)(\partial_{\rm V,T,T})}{2(\partial_{\rm V}P)^2}.
\label{A-15}
\end{equation}
According to Eqs. (\ref{3-0-cc}) and (\ref{A-3}), the normalized scalar curvature of the EHAdS BH with high-order QED correction situation can be written as

\begin{eqnarray}
R_{\rm N}=&&\frac{(1-3\nu'^2)(1-3\nu'^2+4\nu'^3\tau')}{2(1-3\nu'^2+2\nu'^3\tau')^2}\nonumber\\
&&+\frac{a(-2\nu'^2-\nu'^6+3\nu'^8)\tau'^2}{36Q^2(1-3\nu'^2+2\nu'^3\tau')^3}\nonumber\\
&&-\frac{a^2(-2-\nu'^4+3\nu'^6)(-6-\nu'^4+3\nu'^6+4\nu'^7\tau')\tau'^2}{10368\nu'^2Q^4(1-3\nu'^2+2\nu'^3\tau')^4}\nonumber\\
&&-\frac{\beta a^2(-2-\nu'^8+3\nu'^{10})\tau'^2}{864\nu'^2Q^4(1-3\nu'^2+2\nu'^3\tau')^3}+\mathcal{O}(a^3).
\end{eqnarray}

\begin{figure*}[htbp]
  \centering
  \includegraphics[width=7.5cm,height=5cm]{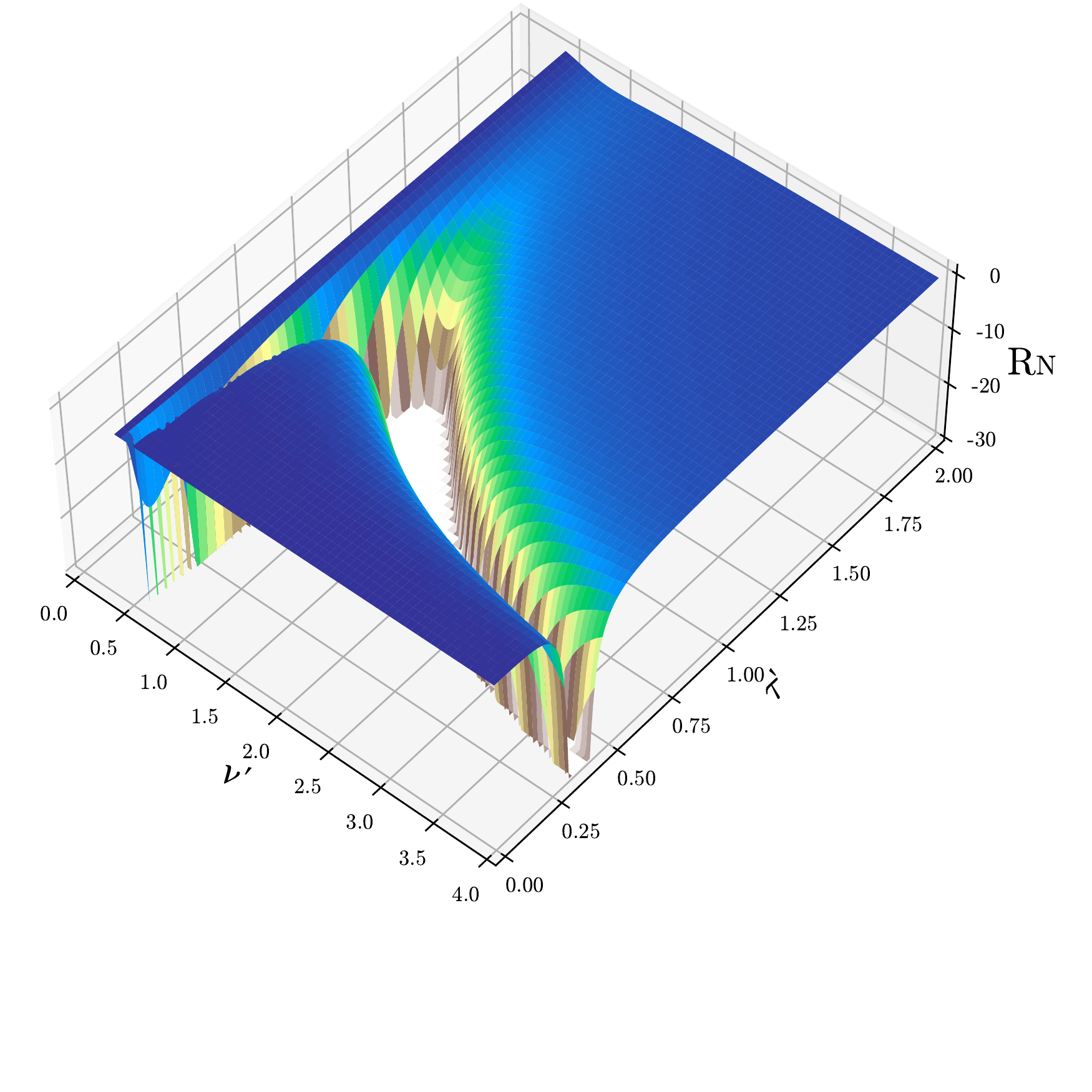}
  \hspace{0.1cm}
  \includegraphics[width=7.5cm,height=5cm]{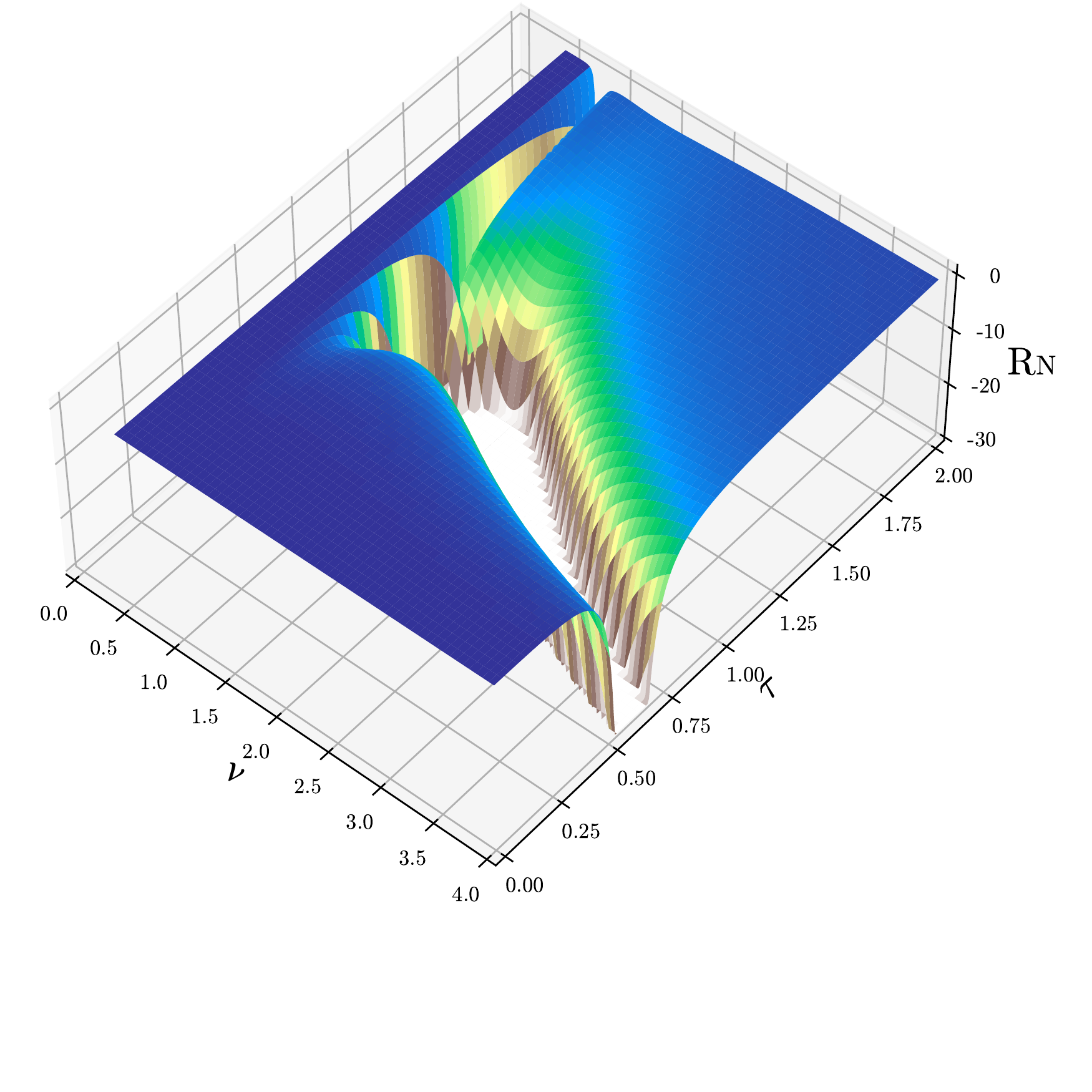}
  \caption {Illustration of the normalized scalar curvature $R_{\rm N}$ as a function of $\nu$ and $\tau$ by setting the EH parameter $a=1$ and the BH charge $Q=0.6$. The left panel is the corrected EHAdS BH ($\beta=1$) and the right panel is the EHAdS BH.}\label{fig:9}
\end{figure*}

\par
Figure \ref{fig:9} illustrates $R_{\rm N}$ as a function of $\nu$ and $\tau$ for the EHAdS BH with/without high-order QED correction. It is observed that the surface of the normalized scalar curvature is concave where the scalar curvature diverges. There is one concave surface for the corrected situation, which is same as the vdW system. Differently, two concave surfaces occur in the EHAdS BH without correction, indicating that the correction leads to the phase transition instability disappears for the EHAdS BH from the microscopic point of view.

\par
Because the vdW type phase transition shows a charge-independent property in the reduced parameter space, we can use the parametrization form to fit the numerical data of the coexistence curves between $\nu'$ and $\tau'$. The parametrization form is
\begin{eqnarray}
\nu'=\sum_{i=0}^{10}a'_{\rm i}\tau'^{i},~~~~\tau'\in(0,1).
\label{A-17}
\end{eqnarray}
The numerical results of the $a'_{\rm i}$ are listed in \ref{Tab:1}, where $\nu'_{\rm l}$ and $\nu'_{\rm s}$ represent the $\nu'$ in the large and small BH region. Figure 10 shows that $\nu'$ as a function of $\tau'$. The small BH volume increases with the temperature increase, which is the opposite of the large BH volume.

\begin{table*}
\caption{ Values of the coefficients $a'_{\rm i}$ in the fitting formula of the coexistence curves for $a=1$, $Q=0.6$ and $\beta=1$}
\label{Tab:1}
\begin{ruledtabular}
\begin{tabular}{c|ccccccccccc}
  $-$ &${a'_{\rm 0}}$ &${a'_{\rm 1}}$ &${a'_{\rm 2}}$ &${a'_{\rm 3}}$ &${a'_{\rm 4}}$ &${a'_{\rm 5}}$ &${a'_{\rm 6}}$ &${a'_{\rm 7}}$ &${a'_{\rm 8}}$ &${a'_{\rm 9}}$ &${a'_{\rm 10}}$ \\
  \hline
  $\nu_{\rm Ql}$ &$90.66$     &$-1131.11$   &$7625.30$    &$-31455.36$ &$83299.04$   &$-143438.45$ &$157364.01$  &$-102074.48$ &$30636.87$   &$1373.82$   &$-2289.27$ \\
  \hline
  $\nu_{\rm Qs}$ &$0.280$       &$0.141$       &$0.0902$     &$0.0668$      &$0.0548$      &$0.0483$      &$0.0450$      &$0.0434$      &$0.0430$      &$0.0432$     &$0.0439$  \\
\end{tabular}
\end{ruledtabular}
\end{table*}
\begin{figure}[htbp]
  \centering
  \includegraphics[width=7.5cm,height=5cm]{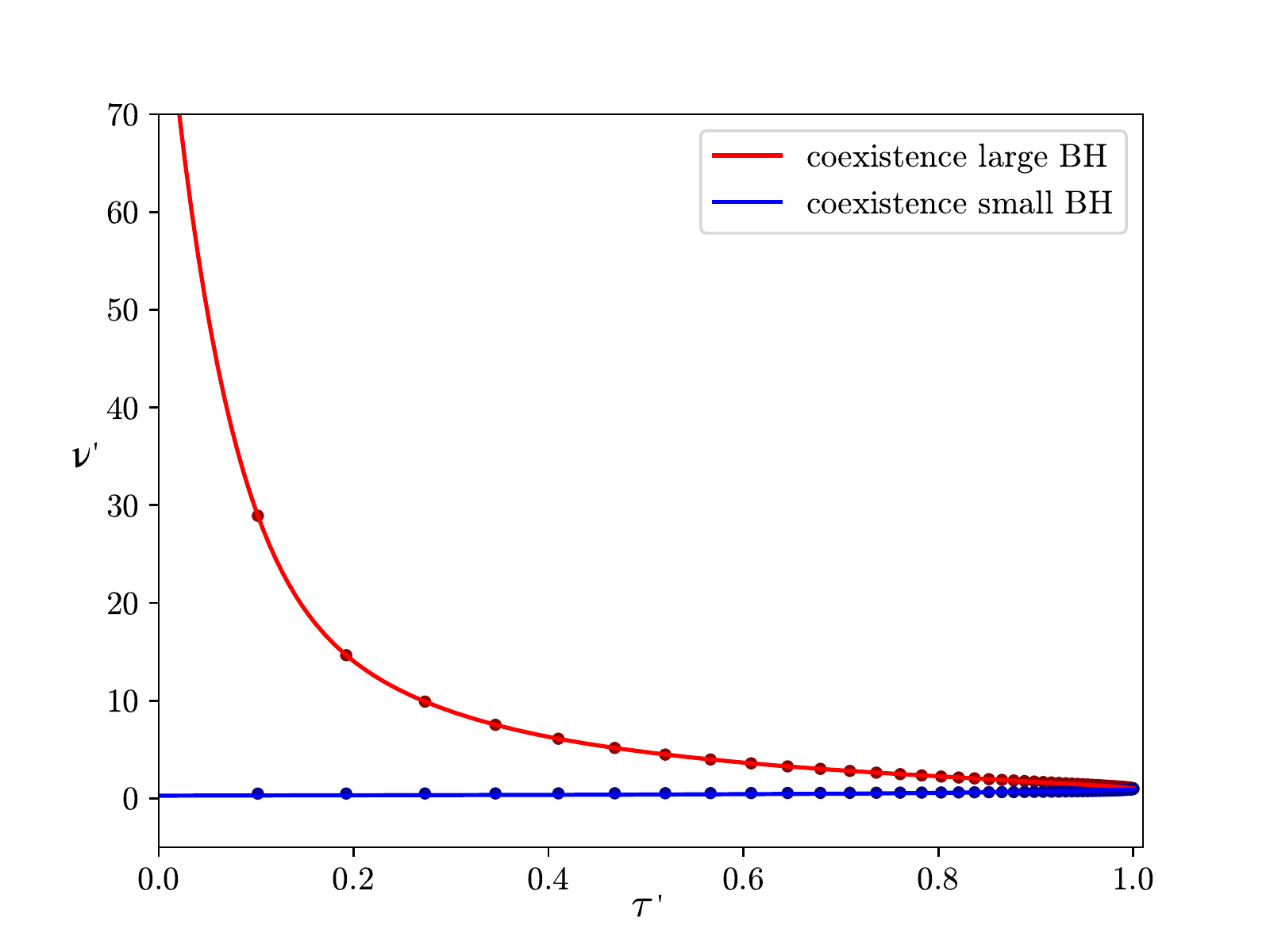}
  \caption {The coexistence curves of $\nu'$ and $\tau'$. The discrete points denote the numerical values and the red/blue line is the fitting formula Eq. (\ref{A-17}). We take $a=1$, $Q=0.6$ and $\beta=1$.}\label{fig:10}
\end{figure}

\begin{figure}[htbp]
  \centering
  \includegraphics[width=7.5cm,height=5cm]{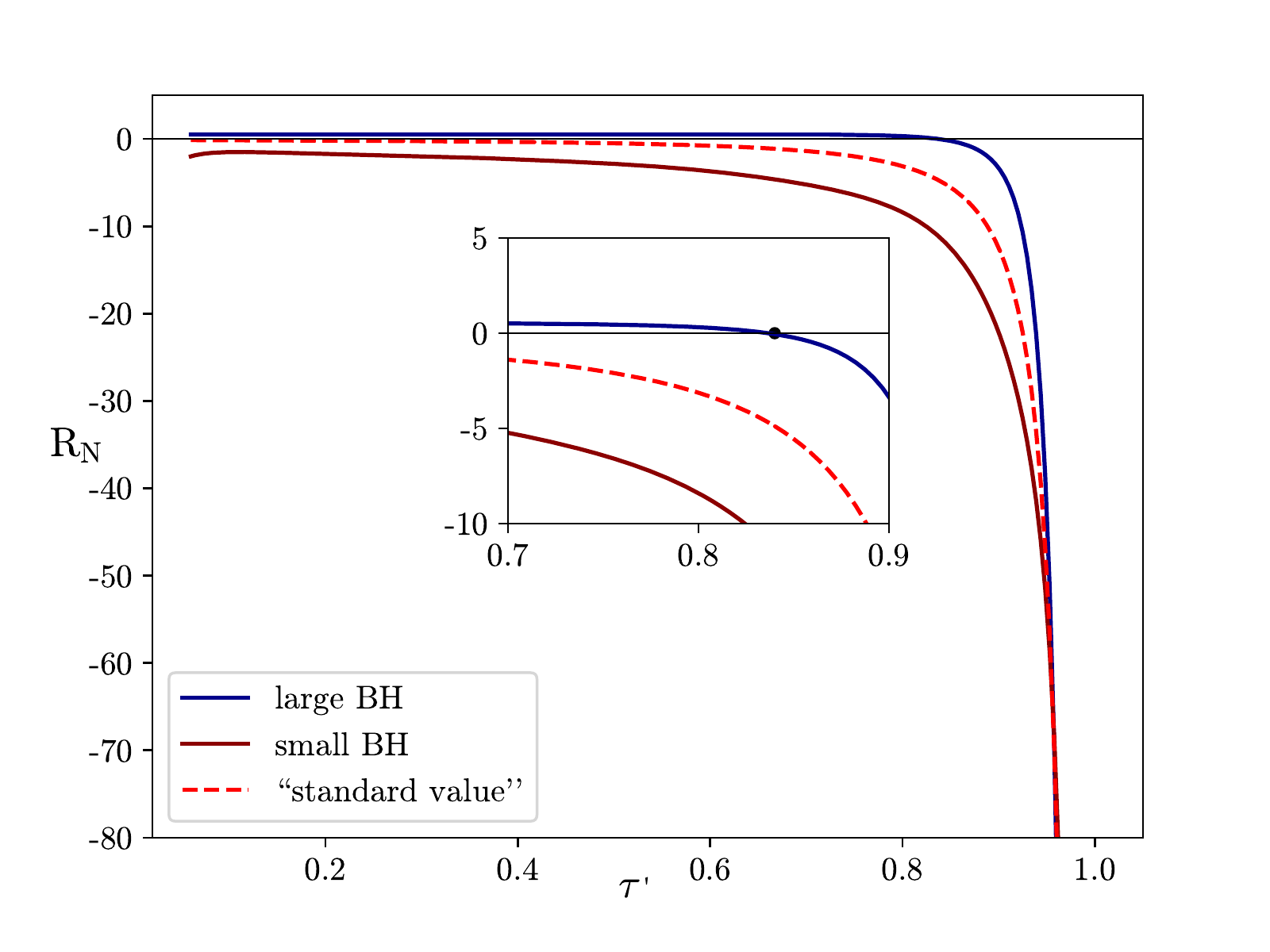}
  \caption {$R_{\rm N}$ along the coexistence saturated large BH (red curve) and small BH (blue curve). The red dashed curve is $R_{\rm N}=-\frac{1}{8}(1-\tau')^{-2}$ as standard property of vdW type phase transition. We take $a=1$, $Q=0.6$ and $\beta=1$.}\label{fig:11}
\end{figure}

Figure \ref{fig:11} shows $R_{\rm N}$ as a function of $\tau'$. We can see that the behavior of the normalized scalar curvature $R_{\rm N}$ along the coexistence of saturated large BH and small BH curves meets the relationship $R_{\rm N}(1-\tau')^2 \sim -\frac{1}{8}$. In the small BH region, $R_{\rm N}$ has a sign change from positive to negative, implying that the dominant micro interaction force transits from repulsion to attraction in the small BH region. These features also appear in charged AdS BH system \cite{9}.

\section{Conclusions}
\label{sec:4}
\par
The high-order QED correction effect on the EHAdS BH phase transition has been revealed in this analysis. Without considering the high-order QED correction, two-phase transition branches and thermodynamic instability are found in the EHAdS BH. By establishing its equal area law, we show that the two-phase transition branches co-exist in the specific temperature and pressure ranges, suggesting that the instability disappears and the reentrant phase transition can occur in this scenario.

\par
Considering the high-order QED correction, we derive the corrected EHAdS BH solution. Only one phase transition branch is found in this scenario by using the critical condition and equation of the state. Its universal constant is $\varepsilon' = 3/8$ for the zeroth-order term of $a$ and the 2nd-order term is 13/82994 in comparison that without the high-order QED correction. Its critical exponents $(\alpha,~\beta,~\gamma,~\delta)$ are $(0,~1/2,~1,~3)$, which is equivalent to the vdW system. This implies that the phase transition of the high-order QED corrected EHAdS BH satisfies the Maxwell behavior. Meanwhile, the sign of the $C_{\rm P'}$ in this scenario does not change, suggesting that the thermodynamic instability disappears for the BH. We also constructed the Maxwells equal area law in this situation and found that it is same as the vdW system and charged AdS BH.

\par
We further investigated the phase transition microstructure of the corrected EHAdS BH through the Ruppeiner geometry. It is found that the surface of the normalized scalar curvature is concave where the scalar curvature diverges. There is one concave surface for the high-order QED correction situation, which is the same as the vdW system. Compared with this scenario, two concave surfaces occur in the EHAdS BH without high-order QED correction, indicating that the correction leads to the phase transition instability disappearing from the microscopic point of view. From the critical behavior of the normalized scalar curvature, it can be further found that the micro characteristics of the corrected EHAdS BH are no different from that of AdS charged. These results suggest that the first phase transition branch of the EHAdS BH is in a metastable state, and its thermodynamics is unstable. The high-order QED correction eliminates this instability, and the phase transition of the EHAdS BH under this correction satisfies the Maxwell behavior.

\begin{acknowledgments}
This work is supported by the National Natural Science Foundation of China (Grant No. 12133003).
\end{acknowledgments}

\end{CJK}
\end{document}